\documentclass{jps-cp}
\usepackage{txfonts} %Please comment out this line unless the txfonts package is availabe in your LaTeX system.
\usepackage{tipa}
\usepackage{tabularx}
\usepackage{makecell}
\usepackage{multirow}
\usepackage{colortbl}
\usepackage{arydshln}
\usepackage{array}
\usepackage{graphicx}

\title{Strangeon Stars}

\author{Jiguang \textsc{Lu}$^{1,2,3}$ and Renxin \textsc{Xu}$^{2,3,4}$}

\inst{$^{1}$National Astronomical Observatories, Chinese Academy of Sciences, Beijing 100012, China\\
$^{2}$State Key Laboratory of Nuclear Science \& Technology, Peking University, Beijing 100871, China\\
$^{3}$Department of Astronomy, School of Physics, Peking University, Beijing 100871, China\\
$^{4}$Kavli Institute for Astronomy and Astrophysics, Peking University, Beijing 100871, China}

\email{lujig@pku.edu.cn, r.x.xu@pku.edu.cn}

\recdate{May 31, 2017}

\abst{
Stable micro-nucleus is 2-flavored ($u$ and $d$), whereas stable macro-nucleus could be 3-flavored ($u$, $d$ and $s$) if the light flavor symmetry restores there. Nucleons are the constituent of a nucleus, while strangeons are named as the constituent of 3-flavored baryonic matter.
Gravity-compressed baryonic object created after core-collapse supernova could be strangeon star if the energy scale ($\sim 0.5$ GeV) cannot be high enough for quark deconfinement and if there occurs 3-flavor symmetry restoration.
Strangeon stars are explained here, including their formation and manifestation/identification.
Much work, coupled with effective {\em micro-model} of strangeon matter, is needed to take advantage of the unique opportunities advanced facilities will provide.}

\kword{pulsar: general, dense matter, equation of state}

\begin{document}
\maketitle

Although the state equation of dense matter around nuclear density is a great challenge for physicists, it is usually a thought that {\em strangeness} could be relevant to the solution.
As noted in the QCS2014 proceedings~\cite{xu15}, such kind of strange matter is conjectured to be strangeon matter, being manifested in the form of compact stars, cosmic rays, and even dark matter.
In this contribution, we are focusing on strangeon star, which is a key ingredient of strangeon matter in astrophysics.

It is half a century since the first pulsar was discovered, but the real nature of pulsar remains one of the most puzzling problems both in physics and in astronomy.
Pulsars are always thought to be neutron stars, but the issue of their real structure is still controversial.
Different observations (e.g., the spin period, the measurements of mass and radius) show that the typical density of pulsar-like compact stars could be only a few nuclear density, and the separation between quarks is $\sim 0.5$ fm and hence the energy scale is order of $\sim 0.5$ GeV according to Heisenberg's uncertainty relation.
The pulsar structure is then a problem of non-perturbative QCD (quantum chromo-dynamics), an unsolved issue for fundamental strong interaction at low energy regime ($\lesssim$ 1\,GeV).
Similar to the issue of turbulence, the strong interaction physics at low-energy is still challenging though QCD is mathematically well-defined.
In fact, both of ``Navier-Stokes Equation'' and ``Yang-Mills and Mass Gap'' are prize problems named by the Clay Mathematical Institute.
Hitherto now, no one can derive pulsar structure from ab-initio calculations, nonetheless astrophysicists could speculate phenomenologically.

\begin{figure}[tbh]
\centering
\includegraphics[width=12cm]{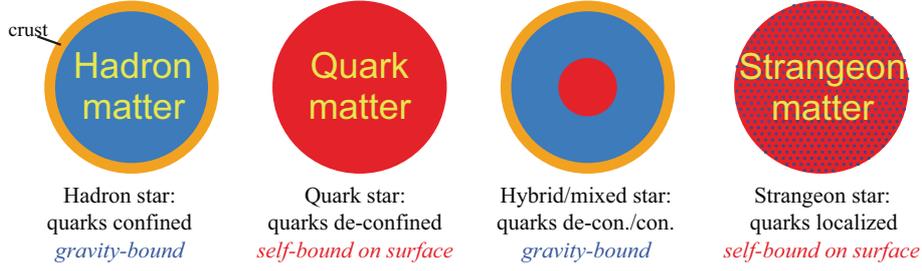}
\caption{Different models of pulsar inner structure.
Hadron star and hybrid/mixed star belong to conventional neutron stars, while strangeness plays an important role in strange quark star and strangeon star as a result of three light-flavour ($u$, $d$ and $s$) symmetry restoration.
}
\label{f1}
\end{figure}
In Fig.~\ref{f1}, we summarize pulsar inner structures (i.e., models) speculators suggest. All of the pulsar structure models fell into four categories: hadron star, quark star, hybrid/mixed star and strangeon star, as are shown in Fig.~\ref{f1}.
In the hadron star model, quarks are confined in hadrons, and the
hadron liquid is the dominated part of the star.
Hadron stars are bounded by gravity, and they must have crusts composed of ions and electrons. A hadron star is called a neutron star if only the nucleon (proton and neutron) degree of freedom is introduced.
Conversely, a quark star is composed of free quarks.
Quarks are bounded by strong-interaction on stellar surface, therefore a quark star can be ``bare'', i.e., a sharp density change exists, from a few nuclear density to zero on the surface.
Although the asymptotic freedom nature is well known for QCD at high energy limit, whether the energy scale of pulsar is high enough for a phase transition from hadron matter to quark matter is still unclear.
Obviously the density/energy scale increases from stellar surface to core, then the hybrid/mixed star model is possible if hadron-quark phase transition occurs inside pulsar.
Hadron matter and quark matter might coexist and crust is also necessary in this model.
Besides, pulsar could be strangeon star (astrophysical condensed-matter of strangeons) if the density is not high enough for quark deconfinement and if 3-flavor symmetry restores in order to make a lower Fermi energy of dense electron gas.

\section{What is a strangeon?}

Strangeon \textipa{["streIdZI6n]} is a new word, formerly known as ¡°strange quark-cluster¡±~\cite{xu03,lai09,guo14}, and is actually coined by combining ``strange nucleon''.
This word is created to describe a new kind of constituent similar to nucleon but with strangeness, being also regarded as an abbreviation of strange nucleon.
The idea of strangeon is related to Witten's conjecture~\cite{witten84}, in which bulk strange {\em quark} matter (i.e., with 3 flavor symmetry of {\em free} $u$, $d$ and $s$ quarks) is speculated to be the ground state of strong-interaction matter.
Pulsar should be strange quark star if this conjecture is correct.
Being very different from strange quark matter, in strangeon matter, a few quarks are localized in a cluster called strangeon if quark-deconfinement energy scale cannot be reached inside pulsars.
A strangeon may contain 6, 9, 12, 18, or even more quarks, but the real number is not determined still.
It is well known that is $\mathrm{\Lambda}$ ($uds$) is 3-flavoured baryon, but the results of both bag model and lattice QCD calculations show evidence for attraction between two $\mathrm{\Lambda}$'s and thus for the existence of a bound H-dibaryon with valence quark structure {\em uuddss}~\cite{jaff77,bean11,inou11}.
The isospin quantum number of strangeon is usually zero.

What's the difference between strangeon and hyperon?
Yes, hyperon is strange baryon, but not the same of strangeon.
First, the number of valence quarks inside a hyperon is 3, while that inside a strangeon is at least 6. The strangeness of a strangeon is usually $S=-B$, with the baryon number $B>1$.
Second, strangeon could be stable in {\em macro}-condensed matter, as an analogy of stable nucleon in a {\em micro}-nucleus, since things may change when the length scale increases (i.e.,  ``bigger is different'').
Third, in vacuum, a hyperon will decay relative slowly via weak interaction, whereas a strangeon may decay quickly into hyperon via strong force. It is well known that neutrons are long-lived inside a nucleus against $\beta$-decay, strangeons as well could be long-lived too in form of condensed matter, to be standing against both strong and weak decays.

In fact, a strangeon star can also be regarded as a macro-nucleus.
Both of normal micro-nucleus and macro-strangeon matter are self-bound by the strong force.
Consequently, one may expect that the interaction between stranegons could be Yukawa-like, and the equation of state (EoS) of Lennard-Jones strangeon matter would then be obtained in a corresponding state approach~\cite{guo14}.

\section{Strangeon matter formation}

De-electronization (i.e., to kill electrons) is essential to set the state of gravity-compressed baryonic matter in stellar collapse.
\begin{figure}[htb]
\centering
\includegraphics[width=11.2cm]{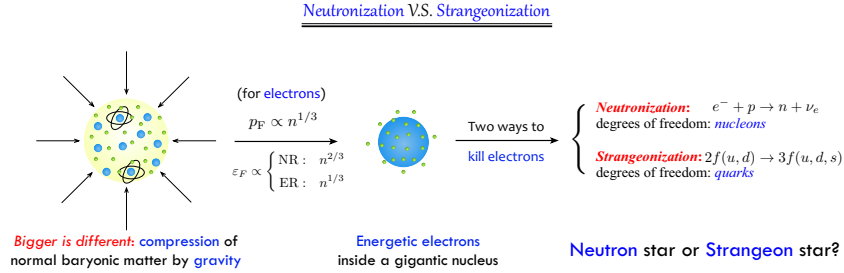}
\caption{A comparison between neutronization and strangeonization, that are two scenarios proposed to kill electrons (i.e., to de-electronize) in core collapse process of a massive evolved star.
}
\label{f2}
\end{figure}
Neutronization is popularly suggested to be the way of de-electronization from Landau's era, however, strangeonization would be an alternative one. It is worth noting that both neutronization and strangeonization are through weak interaction.
Electrons contribute negligible energy in normal {\em electromagnetic} matter (i.e., atoms and ions), but become extremely relativistic when all of the collapsing-core nuclei touch and form a gigantic nucleus (i.e., {\em strong} matter). Bigger is then different: micro-nucleus is one thing, but macro-nucleus is another.
Two candidate scenarios (neutronization and strangeonization) to de-leptonize those energetic electrons are compared in Fig.~\ref{f2}, and consequently two models (neutron star and strangeon star) are proposed.
The former is based on nucleon degree of freedom, while the latter on quark degree.

Normal nucleus could be converted into strangeon matter via the weak interaction during an accretion phase of a strangeon star.
Certainly, a flavor change could be successful only after frequent collisions between a falling nucleus and the strangeon star, and a term of
strangeness barrier is then introduced~\cite{xu14}.
Normal 2-favored baryonic matter blocked by the strangeness barrier wold reside above the strangeon matter surface, forming a corona, an atmosphere or even a crust (being dependent on accretion rate).
A conora/atmosphere would be necessary to understand a low red-shifted O VIII Ly-$\alpha$ emission line~\cite{xu14} and the optical/UV excess of X-ray-dim isolated neutron stars (XDINSs)~\cite{wang17}.

It is possible to synthesize strangeon matter during the cosmic separation of QCD phases.
When compared with strange quark nugget, strangeon nugget with stronger interaction (quarks can then be localized inside strangeon) could have survived in the present Universe, to be manifested in the form of dark matter, since the effects of bolling and evaporation could be weaker.
Binary merger of strangeon star might also eject relativistic strangeon nuggets detectable as cosmic rays.

\section{Strangeon star identification}

It is surely important to identify a strangeon star in order to understand the state of matter at supra-nuclear density and thus the nature of non-perturbative strong force.
The peculiar features of strangeon stars are summarized in Table~\ref{t1}, that could be necessary to understand various observations related to both the {\em surface} condition and the {\em global} structure of strangeon stars.
\begin{table}[tbh]
\caption{To understand different manifestations of compact stars in strangeon star model}
\label{t1}
\begin{small}
\begin{tabular}{cc|c|c|c}
\Xhline{1pt}
\multirow{7}{*}{\rotatebox{90}{\large{surface\qquad\quad}}}    &   Peculiarity &
Manifestation   &   Mechanism & Ref.    \\
\hline
    &   \multirow{2}{*}{binding energy} &   drifting subpulse, micro-structure  &
gap sparking    &   \cite{xu98,yu11} \\
\cdashline{3-5}[0.8pt/2pt]
    &   &   clean fireball for SNE/SGR  &   photon-driven explosion &
\cite{chen07,dai11}   \\
\cdashline{2-5}[0.8pt/2pt]
    &   self-bound  &   mass as low as $\sim10^{-2}M_{\odot}$   &
bound not by gravity    &   \cite{xu03a,xu05}   \\
\cdashline{2-5}[0.8pt/2pt]
    &   \multirow{2}{*}{non-atomic X-rays}   &   Plankian radiation of X-ray    &
very thin atmosphere above surface   &   \cite{xu02,wang17}   \\
\cdashline{3-5}[0.8pt/2pt]
    &   &   absorption in thermal spectra &  hydromagnetic oscillation of $e^-$ sea  &
\cite{xu12} \\
\cdashline{2-5}[0.8pt/2pt]
    &   \multirow{2}{*}{strangeness barrier}    &   low-z emission, type-I XRB  &
2-flavored matter separated from 3-{\em f} &   \cite{xu14}   \\
\cdashline{3-5}[0.8pt/2pt]
    &   &   optical/UV excess of XDINS   &   bremsstrahlung radiation    &
\cite{wang17}  \\
\hline
\multirow{3}{*}{\rotatebox{90}{\large{global~}}}    &   stiff EoS   &   high $M_{\rm max}$ ($2\sim3M_{\odot}$)  &
non-relativistic strangeon, hard core    &   \cite{lai09a,lai09b,lai09,guo14}  \\
\cdashline{2-5}[0.8pt/2pt]
    &   anisotropic pressure   &   SGR/AXP¡¯s burst and flare   &   quake-induced energy release   &   \cite{xu06a,zhou14,lin15}  \\
\cdashline{2-5}[0.8pt/2pt]
    &   rigidity    &   precession, GW radiation    &
solid, mountain building    &   \cite{xu03,xu06b}   \\
\Xhline{1pt}
\end{tabular}
\end{small}
\end{table}

A conventional neutron star has to have a crust due to gravity-induced chemical equilibrium, while a strangeon star could be bare (i.e., nothing outside the surface of strangeon matter) or covered by a corona/atmostpher/crust because of the strangeness barrier. It is reasonable that electromagnetic matter should be separated sharply from strong matter.
Bare surface of strangeon star could help in understanding the non-atomic spectra of XDINSs~\cite{xu02}.
Additionally, the high binding energy in strangeon matter surface would be necessary for polar gap sparking in pulsar magnetosphere~\cite{xu98,yu11}, being essential to understand radio drifting subpulses (even bi-drifting subpulses~\cite{qiao04}) detected, but the binding energy of surface neutron star matter can not high enough to work~\cite{mull84,jone86}.

As gravity-bound body, neutron star's radius decreases as mass increases, and the minimum mass could be order of $0.1M_{\odot}$.
The EoS of hadron star softens if the hyperon degree and even free quarks are included.
However, strangeon star is self-bound, then its radius increases almost with mass, and its mass cold be as low as planet-mass (i.e., strangeon planet~\cite{xu03a}).
Because of the massive (thus non-relativistic) nature of strangeons and the short-distance repulsion force between strangeons (an analogy of the nuclear hard core), the EoS of strangeon matter is very stiff so that the observations of massive pulsars are expected~\cite{lai09}, and the maximum mass can even approach $3M_{\odot}$.
Pulsars as massive as $2M_\odot$, and even with masses higher than $2.5M_\odot$ could then be natural in the strangeon star model.
In addition to the maximum mass, the different mass-radius relations would also be helpful to distinguish between strangeon star and neutron star models.
In fact, the photospheric radius expansion bursts of 4U 1746-37 show that the compact object could be a low mass strangeon star candidate~\cite{li15}.
A strangeon star is globally solid, whose spin would not be limited by the $r$-mode instability and gravitational wave radiation (especially for low-mass strangeon star)~\cite{xu05}, pulsars with rotation periods shorter than 1.3 millisecond and even a sub-millisecond spin could exist in principle. Therefore, a discovery of sub-millisecond pulsar would be strong evidence for strangeon star.
Both types of glitches with (type II) and without (type I) X-ray enhancement could be naturally understood in the starquake model of solid strangeon stars~\cite{zhou14}.
Besides, one may predict precession behaviours of strangeon stars since a rigid body usually precesses when spinning, either free or torqued.

In the future, we are trying to identify strangeon stars with some advanced facilities.
For the moment, the largest single dish, Five-hundred-metres Aperture Spherical Telescope (FAST)
has been constructed, and it would provide plenty of new data about radio pulsars.
The Square Kilometres Array (SKA) will be completed in a few years,
and one of its key sciences is related to radio pulsars too.
Moreover, enhanced X-ray Timing and Polarization (eXTP) satellite is proposed and expected to launch around 2025, and it is one of the key sciences to know the EoS of dense matter inside compact stars.
With these facilities, a lot of data about pulsar-like compact stars will be accumulated, and we are expecting to test the strangeon star model further.

\section{Conclusion and Outlook}

An extension of Witten's conjecture suggests that strangeon matter in bulk
may constitute the true ground state of strong-interaction matter,
and pulsar-like compact stars could then be strangeon stars (i.e., objects composed by condensed matter of strangeons). A lump of strangeon matter could be regarded as a gigantic nucleus where three-flavor symmetry of quarks ($u$, $d$ and $s$) restores, while a normal micro-nucleus is two-flavored. Strangeonization may occur during core-collapse supernova, converting all of the two-flavored micro-nuclei into a single three-flavored gigantic nucleus (note that Landau's gigantic nucleus is still two-flavored).
Different manifestations would be understood if pulsars are strangeon stars,
and we are expecting to test further by advanced facilities.

Despite that astro-phenomenological studies may support the extension, however, the micro-physical foundation of the strangeon matter conjectured is still not solid.
An effective micro-model of strangeon matter is then really welcome.
It is well known that the relativistic mean field theory is successful for 2-flavored nucleon matter in nuclear physics, and an analogy of this traditional theory but for 3-flavored strangeon matter could be interesting in astrophysics.
This kind of model is not only meaningful for one to understand the micro-physics, but also useful to observationally identify strangeon matter.
According to such a model, one may calculate the probability of normal matter being converted into strangeon matte during accretion, and may even present theoretically an air-shower scenario to guide experimentalists discovering strangeon cosmic ray.
A strangeon cosmic ray can only be identified after its air-shower event is modelled, for instance, in IceCube, Pierre Auger  Observatory, and even Chinese LHAASO under construction.

\vspace{0.6cm}
\noindent
{\bf Acknowledgments:}
We are grateful to the members at the pulsar group of Peking University.
This work is supported by the National Natural Science Foundation of China (11673002 and U1531243) and the Strategic Priority Research Program of CAS (No. XDB23010200).
The FAST FELLOWSHIP is supported by Special Funding for Advanced Users, budgeted and
administrated by Center for Astronomical Mega-Science, Chinese Academy of Sciences (CAMS).


\begin{thebibliography}{9}
\bibitem{xu15} R. Xu, Acta Astronomica Sinica, \textbf{56} ({\em supplement}), 82 (2015)
\bibitem{xu03} R. Xu, ApJL, \textbf{596}, L59 (2003)
\bibitem{lai09} X. Y. Lai and R. X. Xu, MNRAS, \textbf{398}, L31 (2009)
\bibitem{guo14} Y. J. Guo, X. Y. Lai, R. X. Xu, Chin. Phys. C. \textbf{38}, 055101 (2014)
\bibitem{witten84} E. Witten, Phy. Rev., \textbf{D30}, 272 (1984)
\bibitem{jaff77} R.~L.  Jaffe, Phys. Rev. Lett., \textbf{38}, 195 (1977)
\bibitem{bean11} S. R. Beane, (NPLQCD Collaboration) et al., Phys. Rev. Lett., \textbf{106}, 162001 (2011)
\bibitem{inou11}  T. Inoue, (HAL  QCD  Collaboration) et  al., Phys. Rev. Lett., \textbf{106}, 162002 (2011)
\bibitem{xu14} R. X. Xu, RAA, \textbf{14}, 617 (2014)
\bibitem{wang17} W. Y. Wang, J. G. Lu, H. Tong, et al., ApJ, \textbf{837}, 81 (2017)
\bibitem{xu02} R. X. Xu, ApJ, \textbf{570}, L65 (2002)
\bibitem{xu98} R. X. Xu, G. J. Qiao, Chinese Physics Letters, \textbf{15}, 934 (1998)
\bibitem{yu11} J. Yu and R. X. Xu, MNRAS, \textbf{414}, 489 (2010)
\bibitem{qiao04} G. J. Qiao, K. J. Lee, B. Zhang, et al., ApJ, \textbf{616}, L127 (2004)
\bibitem{mull84} E. M\"{u}ller, A$\&$A, \textbf{130}, 415 (1984)
\bibitem{jone86} P.~B. Jones, MNRAS, \textbf{218}, 477 (1986)
\bibitem{xu03a} R. X. Xu and F. Wu, Chinese Physics Letters, \textbf{20}, 806 (2003)
\bibitem{li15} Z. S. Li, Z. J. Qu, L. Chen, et al., ApJ, \textbf{798}, 56 (2015)
\bibitem{xu05} R. X. Xu, MNRAS, \textbf{356}, 359 (2005)
\bibitem{zhou14} E. P. Zhou, J. G. Lu, H. Tong, and R. X. Xu, MNRAS, \textbf{443}, 2705 (2014)
\bibitem{chen07} A. B. Chen, T. H. Yu and R. X. Xu, ApJ, \textbf{668}, L55 (2007)
\bibitem{dai11} S. Dai, L. X. Li and R. X. Xu, Science China G: Physics, Mechanics $\&$ Astronomy, \textbf{54}, 1541 (2011)
\bibitem{xu12} R. X. Xu, S. I. Bastrukov, F. Weber, et al., Phys. Rev. D, \textbf{85}, 023008 (2012)
\bibitem{lai09a} X. Y. Lai and R. X. Xu, Astroparticle Physics, \textbf{31}, 128 (2009)
\bibitem{lai09b} X. Y. Lai and R. X. Xu, MNRAS, \textbf{398}, L31 (2009)
\bibitem{xu06a} R. X. Xu, X. H. Cui and G. J. Qiao, Chinese Journal of Astronomy and Astrophysics, \textbf{6}, 217 (2006)
\bibitem{lin15} M. X. Lin, R. X. Xu and B. Zhang, ApJ, \textbf{799}, 152 (2015)
\bibitem{xu06b} R. X. Xu, Astroparticle Physics, \textbf{25}, 212 (2006)

\end{thebibliography}
\end{document}